\documentclass[preprint,12pt]{elsarticle}

\usepackage{lineno}
\usepackage{ulem}
\usepackage{amsmath,amssymb}
\usepackage{xpatch}
\usepackage{float}
\usepackage{microtype}
\usepackage{afterpage}
\newcommand{\Kra}{$\mathrm{KRA}_\gamma$ }
\newcommand{\Kmin}{$\mathrm{KRA}_\gamma^\mathrm{min}$ }
\newcommand{\Kmax}{$\mathrm{KRA}_\gamma^\mathrm{max}$ }
\newcommand{\Kold}{$\mathrm{KRA}_\gamma^{5\,\mathrm{PeV}}$ }

\usepackage[a4paper, , top=20mm, bottom=20mm, includeheadfoot]{geometry}
\usepackage{soul}
\usepackage{booktabs}
\usepackage{amsmath}
\usepackage{multirow}
\usepackage{graphicx}

\usepackage{etoolbox}
\appto{\bibsetup}{\sloppy\setlength{\emergencystretch}{3em}}

\usepackage{hyperref}                  
\hypersetup{
  colorlinks=true,
  allcolors={red},
  menucolor={red},
}

\title{Search for Diffuse Galactic Neutrinos with the Full ANTARES Telescope Dataset} 

\begin{document}

\fntext[cor1]{T. Cartraud is the corresponding author}

\author[]{\textbf{ANTARES Collaboration}}  

\author[IPHC,UHA]{A.~Albert}
\author[IFIC]{S.~Alves}
\author[UPC]{M.~Andr\'e}
\author[UPV]{M.~Ardid}
\author[UPV]{S.~Ardid}
\author[CPPM]{J.-J.~Aubert}
\author[APC]{J.~Aublin}
\author[APC]{B.~Baret}
\author[LAM]{S.~Basa}
% ...

\address[IPHC]{\scriptsize{Universit\'e de Strasbourg, CNRS,  IPHC UMR 7178, F-67000 Strasbourg, France}}
\address[UHA]{\scriptsize Universit\'e de Haute Alsace, F-68100 Mulhouse, France}
\address[IFIC]{\scriptsize{IFIC - Instituto de F\'isica Corpuscular (CSIC - Universitat de Val\`encia) c/ Catedr\'atico Jos\'e Beltr\'an, 2 E-46980 Paterna, Valencia, Spain}}
\address[UPC]{\scriptsize{Technical University of Catalonia, Laboratory of Applied Bioacoustics, Rambla Exposici\'o, 08800 Vilanova i la Geltr\'u, Barcelona, Spain}}
\address[UPV]{\scriptsize{Institut d'Investigaci\'o per a la Gesti\'o Integrada de les Zones Costaneres (IGIC) - Universitat Polit\`ecnica de Val\`encia. C/  Paranimf 1, 46730 Gandia, Spain}}
\address[CPPM]{\scriptsize{Aix Marseille Univ, CNRS/IN2P3, CPPM, Marseille, France}}
\address[APC]{\scriptsize{Universit\'e Paris Cit\'e, CNRS, Astroparticule et Cosmologie, F-75013 Paris, France}}
\address[LAM]{\scriptsize{Aix Marseille Univ, CNRS, CNES, LAM, Marseille, France }}
% ...

\begin{abstract}
The diffuse emission of $\gamma$-rays and neutrinos, produced by interactions of cosmic rays with interstellar matter in the Milky Way, provides valuable insights into cosmic ray propagation and Galactic processes. Emission models incorporating different assumptions about cosmic ray diffusion, source distribution, and target gas density are tested using data from neutrino telescopes. In this study, the final all-flavor neutrino dataset, collected over 15 years (2007–2022) by the ANTARES neutrino telescope, is analyzed. A maximum likelihood ratio method built to handle templates of Galactic emission models is employed to evaluate the compatibility of these models with the observed spatial and energy distributions of neutrino events. The results do not yield stringent constraints on the tested models and upper limits on the diffuse neutrino flux are derived, which are compatible with the results obtained by other experiments.
\end{abstract}

\maketitle

\tableofcontents

\section{Introduction}

Neutrino astronomy remains in its early stages. Since the pioneering detection of a diffuse flux of cosmic neutrinos by the IceCube Collaboration \cite{HESE_3years}, and its subsequent confirmation through multiple independent analyses \cite{HESE_75years, diff_track_IC}, only three sources of cosmic neutrinos have been identified: the flaring blazar TXS 0506+056 \cite{TXS}, the Seyfert Galaxy NGC 1068 \cite{NGC}, and, more recently in 2023, the Milky Way \cite{ICScience2023, GalRidge}. This discovery was anticipated by the ANTARES Collaboration in~\cite{GalRidge}.

Cosmic rays (CRs) generated within our Galaxy are believed to propagate through complex trajectories influenced by the Galactic Magnetic Field. During their propagation, interactions with interstellar gas, primarily hydrogen and helium, produce pions: neutral pions decay into $\gamma$-rays, while charged pions decay into neutrinos \cite{Stecker_1979}. Thus, understanding the diffuse Galactic neutrino flux is key for deciphering cosmic ray transport mechanisms and differentiating between hadronic and leptonic processes occurring within the Milky Way.

Despite recent progress, our understanding of the Galactic neutrino flux remains limited. Critical aspects such as the absolute flux, spectral shape, and spatial origin lack stringent constraints. Moreover, no high-energy neutrino sources within the Galaxy have been conclusively identified, and the processes responsible for the observed hardening of the CR spectrum toward the Galactic Center observed by $\gamma$-ray observations remain poorly understood \cite{pi0_2012, KRA2015}.

The phenomenological models that are tested in the present work integrate observations of $\gamma$-rays, cosmic rays, and, more recently, neutrinos as constraints for the CR emission and propagation within the Galaxy. These models also aim to decompose the contributions of different flux components (diffuse, unresolved sources), providing predictions for the Galactic neutrino flux. Testing these models is an essential step toward advancing our understanding of Galactic neutrino emissions. Such efforts rely on hypothesis testing with maximum likelihood ratios and accurate detector response functions that preserve the intricate structure of the models.

This study employs a template-based analysis to investigate Galactic neutrinos using the final dataset from the ANTARES telescope. ANTARES \cite{ANTARES}, was a neutrino observatory comprising 12 detection lines with 25 storeys each, encompassing a total of 885 optical modules, which was situated in the Mediterranean Sea off the coast of Toulon, France. The telescope detected the Cherenkov light induced in the medium by charged particles generated in neutrino interactions. Muon neutrinos ($\nu_\mu$) interacting via charged current (CC) processes produce muons, leading to track-like events, while other neutrino interactions produce particle cascades resulting in shower-like events. The ANTARES telescope offers several advantages over IceCube, including superior angular resolution due to the optical properties of water and a location better suited for observing the central part of the Milky Way with a reduced atmospheric background. Additionally, the ANTARES datasets, particularly those involving shower-like events, extend down to a few hundred GeV energy range, complementing IceCube's capabilities. Physics results from the ANTARES detector are summarized in~\cite{Albert_2025}.

Following a review of the phenomenological models analyzed (Section~\ref{Models}), the present study details firstly the methodology applied to process ANTARES data (Section~\ref{datasamples}). A template likelihood analysis (Section~\ref{dataanalysis}) is then presented to test the different phenomenological models presented and the ANTARES neutrino candidates. Particular attention is given to the construction of probability density functions (PDFs). Results are subsequently discussed in Section~\ref{Results}. Section~\ref{GalRidge} contains an update of the analysis of the Galactic Ridge~\cite{GalRidge}, which consists of an on-off counting model-independent search for neutrinos in the Center of the Milky Way.

\section{Comparison of Models}\label{Models}

Hadronic interactions between high-energy CRs with the interstellar gas and radiation fields 
of the Galaxy are expected to produce the diffuse Galactic neutrino emission. A grid of models of production and transport of CRs in the Galaxy using GALPROP code \cite{GALPROP_2022} have been fitted to $\gamma$-ray observations seen by the Fermi-LAT experiment. The resulting Fermi-LAT model ($ ^S {\rm S}^Z 4^R 20^T 150^C 5$ \cite{pi0_2012}, renamed $\pi^0$ by the IceCube Collaboration) assumes a single power law of spectral index $-2.7$ for neutrinos over the full sky. This model reproduces roughly the behavior of the diffuse Galactic $\gamma$-ray flux up to a few hundreds of GeV. However, it does not predict the observed hardening of CR emission in the Galactic Center \cite{pi0_2012, KRA2015}, which has prompted the development of several phenomenological models to account for this anomaly. 

To compensate for this underpredicted flux in the inner Galaxy, one popular hypothesis investigated consists of no longer assuming homogeneous transport of CRs in the Milky Way. Indeed, a space-dependent transport of CRs is theoretically expected to originate from a higher level of turbulence and larger poloidal component of the magnetic field near the Galactic Center.
\Kra models, such as \Kold \cite{KRA2015} -- which was used as a template in the IceCube and ANTARES joint analysis of the Galactic Plane neutrino emission~\cite{Albert_2018}, as well as in the IceCube discovery paper \cite{ICScience2023} -- and its more recent iterations, \Kmax and \Kmin \cite{KRA2022, KRA2022_2}, also called $\gamma$-optimized models, assume spatially-dependent CR transport, with a diffusion coefficient whose energy-dependence varies as a function of the galactocentric radius\footnote{These are publicly available at \url{https://zenodo.org/records/12802088}}. The model \Kold uses DRAGON code~\cite{Dragoncode} for the transport of Galactic CRs, and the newest iterations of the models \Kmax and KRA$_\gamma^\mathrm{min}$, use its second version DRAGON2 \cite{DRAGON2, DRAGON2_2} and HERMES \cite{HERMES} to compute the $\gamma$-ray and neutrino emissivities and integrate them along the line of sight. DRAGON codes allow for propagation with a diffusion coefficient that varies not only with rigidity but also with the galactocentric radius. It leads to a spatially-dependent neutrino spectrum. These models fit the local CR spectral measurements from the GeV to the PeV and some of them reproduce the newest HAWC~\cite{HAWC} and LHAASO~\cite{LHAASO} $\gamma$-ray diffuse data, predicting a flux significantly higher than in the homogeneous scenario at energies above the TeV~\cite{KRA2022_2,KRA2022_2025}. Quite relevant for this work is the fact that these models naturally reproduce the $\gamma$-ray diffuse emission measured by H.E.S.S. \cite{HESS:2006vje} and Fermi-LAT in the Galactic Ridge region \cite{Gaggero:2017jts}.

The second hypothesis consists of taking into account unresolved sources.
The diffuse Galactic neutrino flux is produced after the CRs 
escape their sources of production and while they propagate in the Milky Way, in the interstellar medium. However these interactions might also happen close to an acceleration site of CRs. The neutrinos hence emitted would be associated to Galactic sources. Due to the lack of statistics, these sources cannot be detected yet, leading to an indistinguishability between the true Galactic diffuse neutrino flux, and an unresolved source component in current neutrino observations. The existence of such an unresolved flux has been supported by TeV $\gamma$-ray observations \cite{Vecchiotti_2022}.

{The model discussed in \cite{Vecchiotti_2022, Vecchiotti_2023}, referred to as the ``DiffUSE'' model, incorporates both CR diffuse component (Diff) and Unresolved Source Emission (USE).
The diffuse neutrino flux (Diff) is obtained by using the GALPROP gas map along with a parametric description of the CR distribution in the Galaxy that allows to implement different assumptions for CR space and energy distribution.
Specifically, case B from~\cite{Vecchiotti_2022} is considered, where the CR spectrum is based on local measurements, while the CR flux normalization at various Galactic positions is determined as the solution to a 3D isotropic diffusion equation, assuming a stationary injection rate from the distribution of Supernova Remnants (SNRs).
A diffusion radius equal to $R=1\,{\rm kpc}$ is adopted, which implies that CRs are confined relatively close to their sources.
The Unresolved Source Emission (USE) is derived from the population study of high-energy $\gamma$-ray sources presented in \cite{Cataldo_2019}. 
In this work, sources are assumed to be distributed as pulsars (or SNRs) in the Galaxy and to have a power law luminosity function with unknown normalization and maximal luminosity\footnote{This type of behavior is commonly assumed for various source classes (see, e.g., \cite{Steppa:2020qwe,Lipari:2024pzo}) and it is naturally expected for fading sources, such as pulsar wind nebulae or TeV halos.}. 
These parameters are then constrained by fitting the longitude, latitude, and integrated flux (from 1 to 100 TeV) of distributions of bright sources included in the H.E.S.S. Galactic Plane Survey.
Finally, in order to obtain the neutrino USE, it is assumed that 20\% of these sources is powered by hadronic interactions, which is the maximum allowed based on IceCube measurements for sources with an energy cutoff of 1 PeV, according to the analysis of \cite{Vecchiotti_2023}.

Finally, the CRINGE model~\cite{CRINGE}, which fits high-energy CR data (AMS-02, DAMPE, IceTop, KASCADE), uses the same unresolved contribution as in the DiffUSE model, but also investigates different sources of uncertainties. It uses DRAGON code with different rigidity breaks in the diffusion coefficient, with no position dependency.

In this paper, the sum of the diffuse and unresolved contributions from the CRINGE model is tested, while for the DiffUSE model, only $20\%$ of the unresolved contribution is added to the diffuse model.
The full sky integrated energy spectrum of the different models is plotted on fig.~\ref{fig:flux_E}. The flux integrated over the Galactic longitude $\ell$, at $\sim 1$ TeV, is plotted on fig.~\ref{fig:flux_l}.

\begin{figure}[ht]
    \centering
    \includegraphics[width=\linewidth]{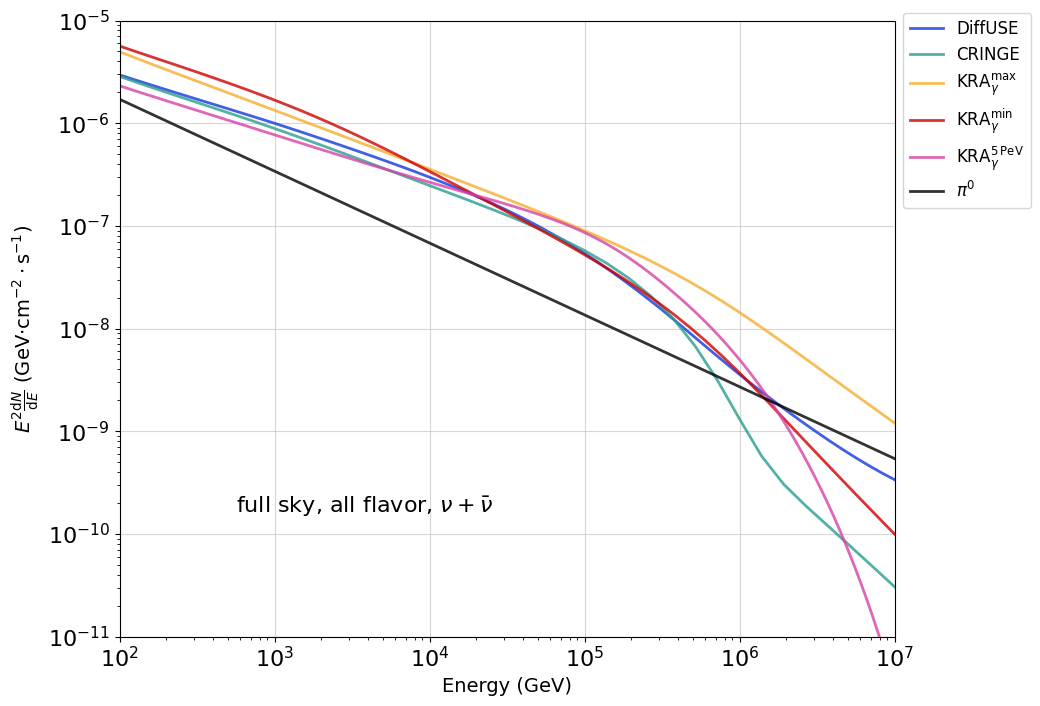}
    \caption{Energy spectrum over the full sky for the different models introduced in the text. The integration is done over all flavors, including both neutrinos and anti neutrinos. While the models predict overall a similar flux, they differ more within specific regions of the Galaxy.}
    \label{fig:flux_E}
\end{figure}

\begin{figure}[ht]
    \centering
    \includegraphics[width=\linewidth]{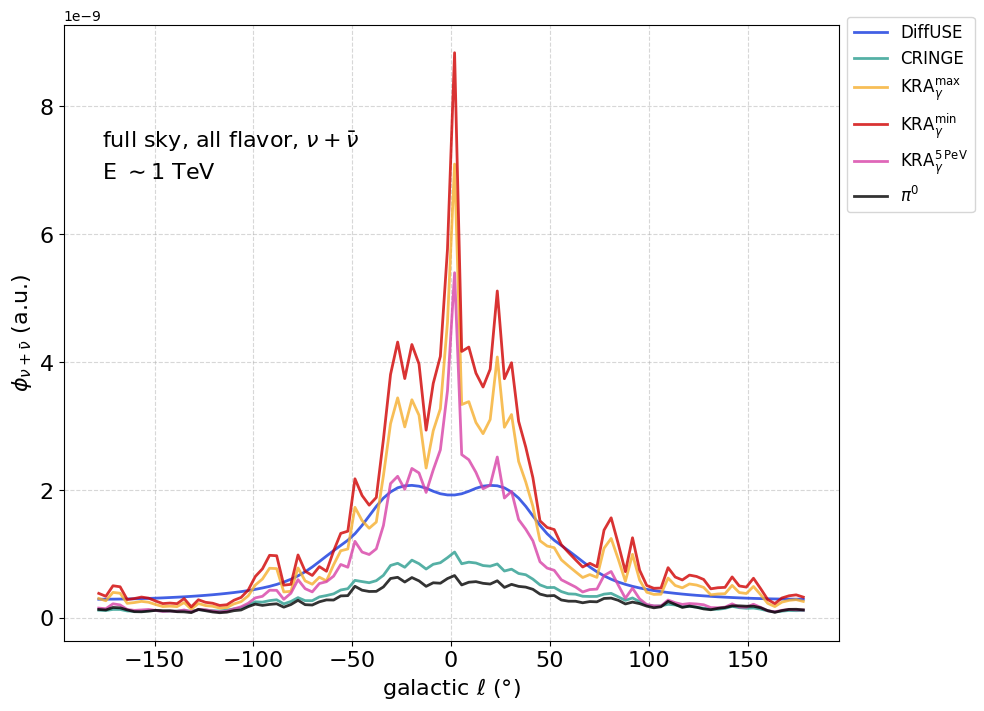}
    \caption{
    Neutrino plus antineutrino all-flavor flux at $\sim 1$ TeV as a function of the Galactic latitude $\ell$  and integrated over Galactic longitude for the different models described in the text. \Kra models are more peaked toward the Galactic Center ($|\ell| < 50^\circ$). DiffUSE model predicts a smooth diffuse flux, as gas targets are modelled using an analytical parametrisation.}
    \label{fig:flux_l}
\end{figure}

\section{Methodologies}\label{Methodologies}

\subsection{Data samples} \label{datasamples}

The ANTARES data and Monte Carlo (MC) simulations used in this analysis correspond to the final and full 15-year sample selected using the criteria already employed for the search of an all-sky diffuse flux of astrophysical neutrinos~\cite{Albert_2024}. Two main topologies of events are investigated, track-like events and shower-like events. The track-like event sample is made of 3392 events. The shower-like events are selected by two sets of cuts, a first one defining a sample of 187 events and a boosted second one, with a statistics of 219 lower-energy events at and below the TeV. They are respectively referred as \emph{shower-high} and \emph{shower-low}, as the second selection is characterized by events at lower energy. All three samples are mutually exclusive and are generated sequentially in the following order: tracks, shower-low, and shower-high. Each selection is optimized based on the remaining events from the previous category.
To reject atmospheric muons, a cut in local zenith at $\cos(\theta) > 0$ on the reconstructed direction of the event is applied to the track-like events to select detected muon entering the detector from below. Other cuts on the reconstruction quality, $\Lambda > -4.9$ (see fig. 2 in~\cite{PRD2017}) and on the estimated angular error in the direction reconstruction, $\beta <0.5^\circ$, which are parameters provided by the track reconstruction algorithm, have both been optimized for the all-sky diffuse analysis~\cite{Albert_2024}.
The main background for shower-like events consists of the hadronic cascades produced by low-energy muons going through a $\nu_\mu$-CC interaction. Due to their similar topologies, additional cuts are indeed required compared to the track-like selection. The shower-like candidates have to be located well inside the detector~\cite{Albert_2024} and to satisfy a likelihood-ratio test criterium between the shower-like and track-like hypotheses~\cite{reco_shower_ANTARES}. A Random Decision Forest (RDF) algorithm~\cite{RDF_shower_ANTARES} is used to discriminate between the neutrino-induced shower-like hypothesis or atmospheric muon hypothesis. Finally, a cut on the angular resolution $\beta < 10^\circ$ is applied.
This procedure is applied for the shower-high sample, and then applied similarly for the shower-low sample. Because shower-like events are lower in energy, they are even more difficult to select, hence a Boosted Decision Tree (BDT) classifier~\cite{BDT_ANTARES} is used as a final step to identify them.
All of the samples are well characterized with a high neutrino purity and show a very robust data/MC agreement. The quality of these event samples has justified the choice to not optimize them further for this analysis. Finally, this sample has an excellent median angular resolution, below 4° for shower-like events and below $0.4^\circ$ for track-like events.

\subsection{The unbinned maximum likelihood analysis} \label{dataanalysis}

An unbinned maximum likelihood ratio test is conducted to determine whether the spatial and spectral distribution of the events is compatible with a background hypothesis $H_b$ or with the additional presence of a signal following a certain hypothesis $H_s$. In this analysis, $m=3$ channels are considered: the track-like and the two different shower-like samples of ANTARES \ref{datasamples}. Each channel $i$th consists of $n_i$ events characterized by their reconstructed positions in equatorial coordinates and reconstructed energies $(\alpha_j^i,\delta_j^i,E_j^i)_{j=1}^{n_j}$ \footnote{the letter $E$ is used in the following to designate $\log_{10}(E/[ \mathrm{GeV}])$.}. The signal and background Probability Density Functions for each channel $i$: $f_s^i$, $f_b^i$ are functions of these coordinates and allow to define the extended log-likelihood functions

\begin{equation}
\begin{split}
        \mathcal{L}_{H_s}(r,\boldsymbol{\mu_b})&= \sum_{i=1}^{m} \Biggl\{ \sum_{j=1}^{n_{i}} \log\Bigl[r \times \mu_{\text{model}}^i \, f_s^i(\alpha_j^i,\delta_j^i,E_j^i)   \\
         &+ \mu_b^i \, f_b^i(\alpha_j^i,\delta_j^i,E_j^i)\Bigr]  - r\times \mu_{\text{model}}^i - \mu_b^i \Biggr\}
\end{split}
\label{eq:loglikelihood_H1}
\end{equation}
\begin{equation}
    \mathcal{L}_{H_b}(\boldsymbol{\mu_b})= \sum_{i=1}^{m}\left\{\sum_{j=1}^{n_{i}} \log\left[\mu_b^i \, f_b^i(\alpha_j^i,\delta_j^i,E_j^i)\right] - \mu_b^i\right\}
    \label{eq:loglikelihood_H0}
\end{equation}
with $m+1$ free parameters, $r$ and $\boldsymbol{\mu_b}$. The vector $\boldsymbol{\mu_b}$ represents the estimated number of background events in each channel sample while the flux ratio $r$ represents the relative model flux normalization (common to all channels). For each channel, the number of signal events $\mu_s^i = r \times \mu_{\text{model}}^i$ can be defined. It depends on $\mu_{\text{model}}^i$, the average number of events expected in this channel for the base value of the flux, i.e.~when $r=1$.
The likelihood functions $\mathcal{L}_{H_s}$ and $\mathcal{L}_{H_b}$ are respectively maximized at the estimators $\left\{\hat{r}, \boldsymbol{\hat{\mu}_b}\right\}$ and $\left\{ \boldsymbol{\hat{\mu}_b}\right\}$, which allow to define the test statistics

\begin{equation}
TS = \text{sign}(\hat{r}) \times \left\{\mathcal{L}_{H_s}(\hat{r},\boldsymbol{\hat{\mu}_b}) - \mathcal{L}_{H_b}(\boldsymbol{\hat{\mu}_b})\right\}.
\end{equation}
The flux ratio estimator $\hat{r}$ is allowed to be negative, which is convenient to detect potential biases in the fit, especially for pseudo-experiments (PEs) with no injected flux (see \ref{PE}). In order to avoid overlaps between the test statistics for positive and negative values of \(\hat{r}\), \(\text{sign}(\hat{r})\) is accounted for in the definition of $TS$.

Minimization is performed using \texttt{iMinuit}~\cite{iminuit}, but issues arise if the flux ratio \(\hat{r}\) is negative. Indeed, in the extended likelihood formalism, the term \(\mu_s f_s + \mu_b f_b\) must remain positive for all events. Allowing negative \(\mu_s\) can make this term negative, rendering the likelihood \(\mathcal{L}_{H_s}\) undefined below certain flux ratios. Near these regions, minimizers may fail to converge. In such cases, \(\mathcal{L}_{H_s}\) is maximized only with respect to $r$, fixing the background events to \(\boldsymbol{\hat{\mu}_b} = \boldsymbol{n_{\mathbf{\text{tot}}}}\).

\subsection{Pseudo-experiments}\label{PE}
Given a hypothesis $H_s$ and a chosen flux ratio $r$, for each channel, $K$ PEs are generated with flux ratios $(r^k)_{k=1}^{K}$ following a normal distribution with a standard deviation of 15\% to account for the systematics coming from the uncertainty on the optical modules acceptance \cite{Adrián-Martínez_2012}. For each of these PEs, the number of detected signal events follows a Poisson law $\mu_s^k \sim \mathcal{P}(r^k \times \mu_\text{model})$. The number of injected background events corresponds to $\mu_b^k = n_\text{tot}-\mu_s^k$ where $n_\text{tot}$ is the number of detected events in the dataset for the given channel. Both PEs following the background hypothesis and the signal hypothesis are generated from the respective PDFs (see \ref{PDF}).
An accurate modeling of the background hypothesis is crucial to avoid biases in the fit of the data. Therefore, TS distributions are generated from data, scrambled in right-ascension. In order to avoid contamination of possible Galactic signal, events located near the Galactic plane and above 10 TeV are removed in the pseudo-experiment generation. The obtained TS distributions match the TS distributions obtained with pseudo-experiments generated from the background PDFs, which ensures an adequate modeling of the background hypothesis and an excellent data/MC agreement.

\subsection{Probability Density Functions} \label{PDF}

\begin{figure}[ht]
    \centering
    \includegraphics[width=0.9\textwidth]{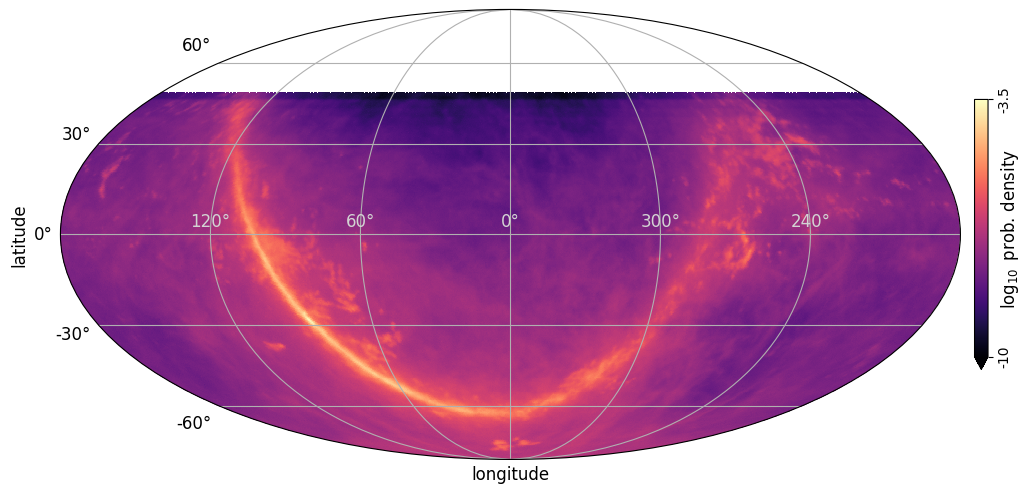}
    \includegraphics[width=0.9\textwidth]{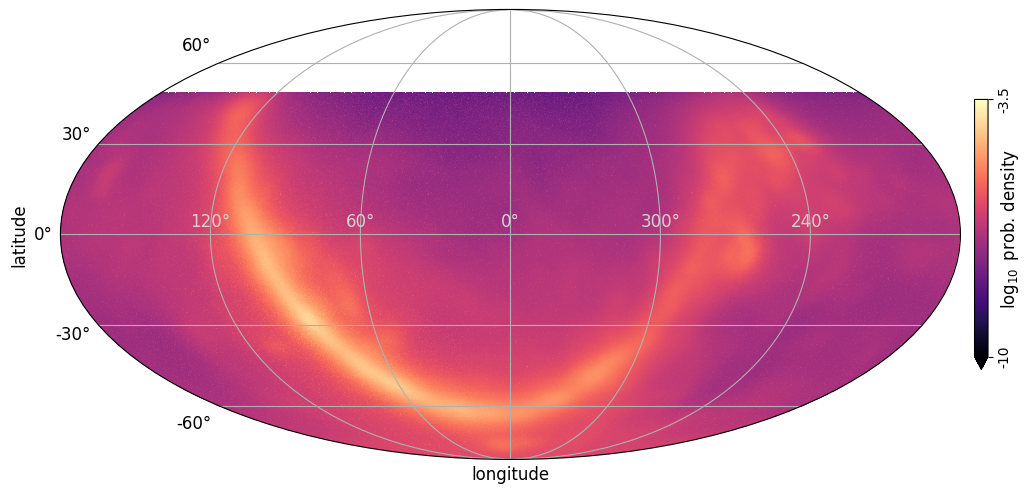}
    \includegraphics[width=0.9\textwidth]{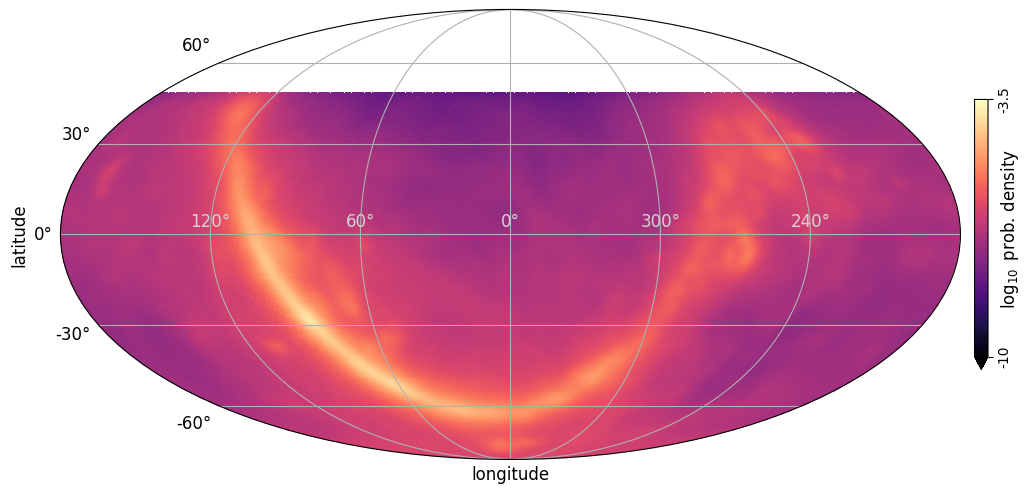}
    \caption{Spatial PDFs of the signal hypothesis as predicted by \Kmax model for the track (top), shower-low (middle) and shower-high (bottom) channels. The predicted neutrino flux is convolved with the ANTARES detector response for each event topology. A logarithmic colored scale is applied going from $-10$ to $-3.5$. The sky-maps are plotted in equatorial coordinates, with a cut in declination to not surpass the ANTARES acceptance limit. The fine spatial structure of the template is preserved, especially for the track channel.}
    \label{fig:background-KDE}
\end{figure}
\afterpage{\clearpage}

The probability density functions $f^i_{s}$ (respectively $f^i_{b})$ represent the probability for any event $j$ of signal (background) origin with true variables $(\alpha_j^{\text{v}},\delta_j^{\text{v}}, E_j^{\text{v}})$ to be detected, selected and reconstructed in the considered channel $i$ as $(\alpha^i_j, \delta^i_j, E^i_j)$.

One traditional way to build PDFs is to factorize $f(\alpha,\delta,E) = g(\alpha,\delta)\times h(E)$ between a spatial and an energy component. However, in the case of the different ANTARES samples tested, the spatial distribution of the events happens to be totally dependent on the energy, making such a factorization impossible. That is why PDFs in this analysis are built as piece-wise functions on the reconstructed energy. To each bin of energy a sky-map is then associated, represented by Healpix maps \cite{Healpix} with a resolution parameter $N_\text{side}=512$ for the track channel and $N_\text{side}=256$ for the shower channels. Bins are defined for every $0.1$ interval in $\log(E^\text{rec}/1\,\mathrm{GeV})$.

The background PDFs appear to be uniform in right ascension $\alpha$ due to Earth's rotation, which washes out possible anisotropies in local coordinates after many years of data taking. A kernel density estimation (KDE) using the SUFTware program \cite{suftware} is hence performed on the $\sin(\delta)$ distribution of the MC events for each energy bin.

Signal PDFs are built by adding to the Healpix sky-maps simulated MC signal events with a weight, that is proportional to the value of the model estimated with the true coordinates and energies $\Phi_\text{model}(\alpha^\text{v}_j,\delta^\text{v}_j,E^\text{v}_j)$. This procedure naturally performs the convolution of the $\nu$ flux model with the ANTARES detector response without the need of any simplified parametrization~(see fig. \ref{fig:background-KDE}).

\subsection{Bootstrap method}~\label{bootstrap}

Signal probability density functions require substantial statistics for accurate modeling. Each pixel of the sky-map must be populated with enough MC events to ensure a smooth flux distribution across the energy range. To enhance statistical accuracy, the following resampling (“bootstrap”) procedure is adopted. MC events are selected within bands of declination in true coordinates, with random positions assigned within each band. Reconstructed coordinates are then randomized by applying a rotation in equatorial coordinates, while maintaining the angular distance between the true and reconstructed positions of each MC event. The randomization in true coordinates along a given declination band is valid due to the uniform detector response in right ascension, and allows to use each MC event in this band as many times as wanted and hence increase the MC statistics. As long as the distances between the new reconstructed and true positions are equal to their progenitors, the detector response characteristics is also preserved.
Since different declination bands may require varying degrees of resampling, the resampling amount is recorded to ensure a correct normalization of the PDF, by not overestimating the weights of events which were resampled more times than others. In regions where the template predicts low flux, statistical under-fluctuations can occur, causing discontinuities in the flux at certain energies. They can be detected thanks to a median filtering on the spectrum derivative. The affected pixels are corrected by taking the median value of the neighbour pixels. A higher amount of resampling also helps to reduce artifacts, but also fluctuations due to the limited amount of resampled events. A final Gaussian smoothing is performed on every map to limit pixel-to-pixel fluctuations, ensuring stable results.

\subsection{Overlapping bins of energy}

Due to the limited amount of MC events, some bins in energy in certain declination bands suffer from a lack of statistics. This issue especially arises when the number of bins in energy increases. The bootstrap method by itself, while addressing this issue, would only resample a few MC events in these regions with poor statistics. To avoid this, a first solution would be to reduce the amount of bins; however, this would lead to a lack of stability of the obtained results.

Indeed, when evaluating signal and background PDFs, the more the bins, the less the weight difference between consecutive bins. As a result, events will have more similar weights when the likelihood is evaluated for adjacent bins, contributing to greater stability in the results. In other words, it is required for the energy distribution of the PDF to be sufficiently smooth.

\begin{figure}[ht]
    \centering
    \includegraphics[width=0.7\linewidth]{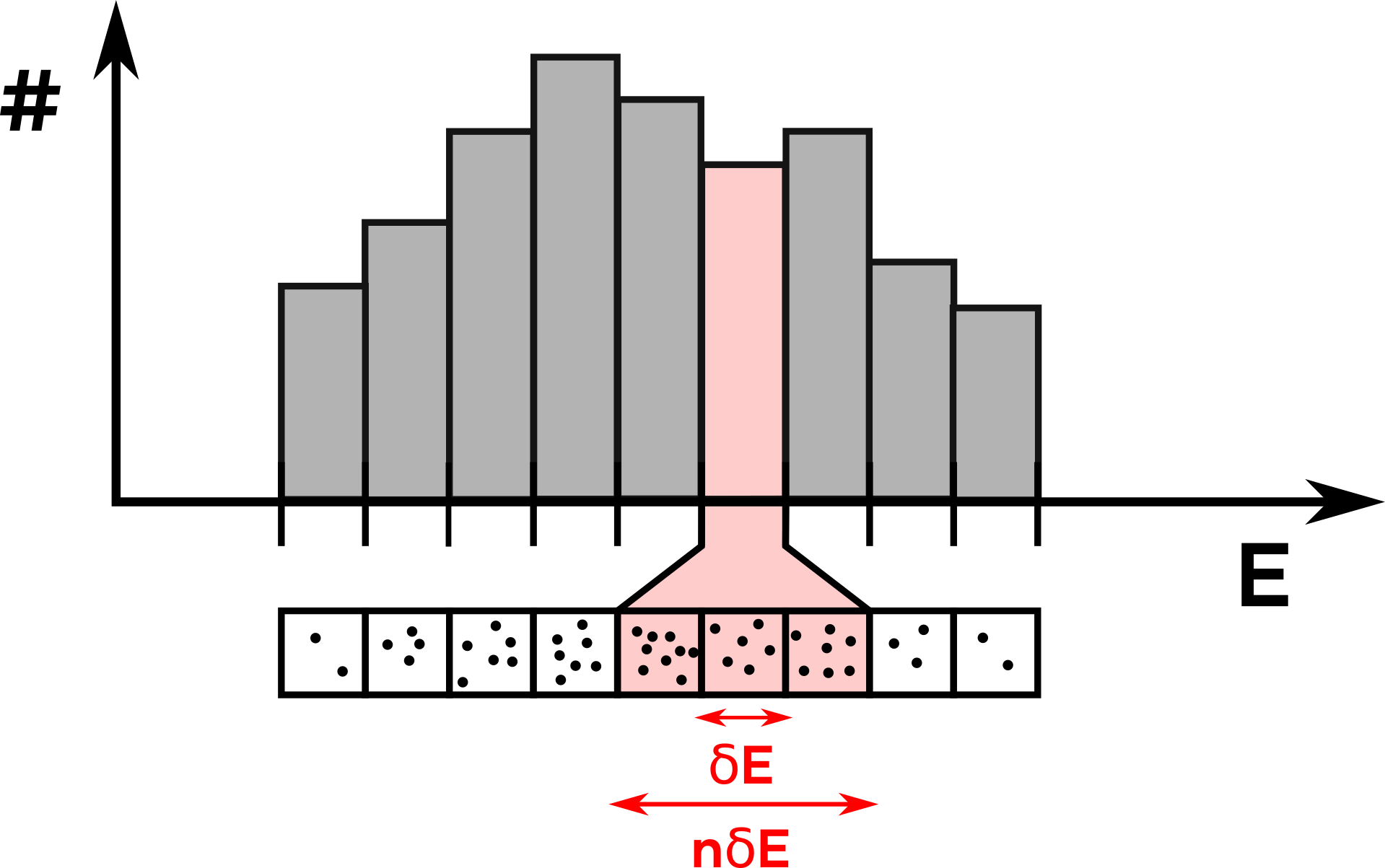}
    \caption{Principle of the overlapping bins of energy. For every bin of size $\delta E$ in energy, the associated spatial PDF is generated using events in the nearby bins (covering a range of $n\delta E$). Taking a fixed amount of nearby bins instead of a generic $\Delta E$ helps to keep track how many times an event is being resampled (see subsec.~\ref{bootstrap}).}
    \label{fig:bins}
\end{figure}

Therefore it is necessary to keep the number of bins in energy high, even if it requires to address the limitations on the statistics. To solve this issue, both background and signal PDFs are built using overlapping bins of energy, which means that, for a given energy bin, events are selected also from outside its edges (see fig.~\ref{fig:bins}). The implementation merges the PDFs of the neighbor bins within a certain range (11 bins every $0.1$ in $\log(E^\text{rec}/1\,\mathrm{GeV})$ into one. The normalization is again ensured so as the weight of a MC event is preserved no matter how many times it is used (see \ref{bootstrap}).
The resulting energy spectrum of the PDFs is similar to a standard energy histogram convolved with a uniform KDE, which is flattened at the top of the distribution and increased at the edges. As it would result in a shape mismatch between the data and the PDF energy spectrum, the weight of every energy bin of the PDF is renormalized to match the standard histogram.

\section{Results from templates}\label{Results}

For each model and its associated \( H_s \) hypothesis, the minimization of the likelihood function for the data yields a fitted test statistic \( \widehat{TS} \), which is used to calculate a p-value and the corresponding one-sided significance. The fitted flux ratio $\hat{r}$ corresponds to the best-fit scaling factor that multiplies the model's predicted flux so the likelihood matches the observed events. The different sensitivities and upper limits are also derived in terms of flux ratios.

Results show (see Table~\ref{tab:merged_sensitivity_flux_upper}) that around half of the flux ratio is fitted for DiffUSE, CRINGE and \Kra models, while a flux ratio close to $3$ is fitted for Fermi-LAT $\pi^0$ model. However, no significant signal is detected as the maximum significance reached is at $1.28\sigma$ for the \Kold model. No significant discrepancy is seen between KRA$_\gamma$ and other models. Therefore, one cannot yet address whether the hardening towards the Galactic Center seen from Fermi-LAT observations~\cite{pi0_2012} is explained by the hardening of the diffusion coefficient towards the Galactic center, and relevant to interpret the physical origins of the Galactic diffuse emission of neutrinos.

\begin{table}[h!]
\centering
\begin{tabular}{|l|c|c|c|c|c|}
\hline
\textbf{Model} & \textbf{$\hat{r}$} & \textbf{$\widehat{T_S}$} & \textbf{p-value} & \textbf{$90\%$ upper limit} & \textbf{Sensitivity} \\ \hline
$\pi^0$ & $2.7$ (9.5 evts) & $0.2$ & $0.22$ ($0.78\sigma$) & $9.4$ (32.5 evts) & $5.7$ (19.8 evts) \\ \hline
KRA$_\gamma^{\text{max}}$ & $0.3$ (6.0 evts) & $0.6$ & $0.11$ ($1.21\sigma$) & $1.2$ (20.6 evts) & $0.6$ (9.5 evts) \\ \hline
KRA$_\gamma^{\text{min}}$ & $0.4$ (6.2 evts) & $0.4$ & $0.15$ ($1.04\sigma$) & $1.4$ (22.3 evts) & $0.8$ (12.2 evts) \\ \hline
CRINGE & $0.6$ (7.5 evts) & $0.2$ & $0.22$ ($0.78\sigma$) & $2.1$ (24.5 evts) & $1.3$ (15.1 evts) \\ \hline
KRA$_\gamma^{5\,\text{PeV}}$ & $0.5$ (6.3 evts) & $0.7$ & $0.10$ ($1.28\sigma$) & $1.4$ (18.9 evts) & $0.7$ (9.8 evts) \\ \hline
DiffUSE & $0.6$ (8.2 evts) & $0.2$ & $0.23$ ($0.73\sigma$) & $2.1$ (27.7 evts) & $1.3$ (17.1 evts) \\ \hline
\end{tabular}
\caption{Summary of results for every model tested. Fitted flux ratios ($\hat{r}$) and fitted test statistics ($\widehat{TS}$) , as well as the p-value and corresponding significance (n$\sigma$) are presented.
The sensitivity, expressed in factors of flux ratios $r$, is best for KRA$_\gamma$ models. Below a flux ratio of 1, the models can be constrained. However, none of the upper limits at a 90\% C.L. are below such ratio. Whenever a quantity is expressed as a flux ratio, the corresponding total amount of ANTARES events is put under parentheses. Note that negative flux ratios, corresponding to negative fitted TS can be obtained, and would have been interpreted as negative fluctuations.}
\label{tab:merged_sensitivity_flux_upper}
\end{table}

\begin{table}[h!]
\centering
\begin{tabular}{|c|c|c|c|}
\hline
\textbf{Model} & \textbf{Track events} & \textbf{Shower-low events} & \textbf{Shower-high events} \\ \hline
DiffUSE & $6.8$ & $1.9$ & $4.3$ \\ \hline
CRINGE & $6.1$ & $1.7$ & $3.8$ \\ \hline
KRA$_\gamma^{\text{max}}$ & $9.2$ & $2.4$ & $5.6$ \\ \hline
KRA$_\gamma^{\text{min}}$ & $8.4$ & $2.7$ & $5.0$ \\ \hline
KRA$_\gamma^{5 \: \text{PeV}}$ & $7.2$ & $1.7$ & $4.5$ \\ \hline
$\pi^0$ & $1.8$ & $0.6$ & $1.1$ \\ \hline
\end{tabular}
\caption{Comparison of the number of signal events predicted by the different models for track, shower-low, and shower-high channels out of the 3392, 219 and 187 events in the corresponding data samples.}
\label{tab:ns}
\end{table}

The sensitivity reported here corresponds to the injected signal flux ratio for which $90\%$ of the test statistics is above the median $H_b$ $TS$ value (see tab.~\ref{tab:merged_sensitivity_flux_upper}). Since the current analysis only fits the model flux normalization, it remains difficult to disentangle the specific impacts of different energy spectra and spatial distributions on the constraints derived for each model. ANTARES energy range extending below TeV energy, especially thanks to the contribution of the shower channels, may help to correlate the amount of constraints on the models to the flux predicted at low energy.

As no significant signal was detected for any of the tested models, 90\% confidence level upper limits using the Neyman construction~\cite{Neyman} are reported in tab.~\ref{tab:merged_sensitivity_flux_upper}. ANTARES dataset allows to derive an excellent sensitivity (below 0.7 for the \Kra models, shown at the last column of tab.~\ref{tab:merged_sensitivity_flux_upper}), however, the resulting bounds do not constrain any models, as the lowest upper limit is at $1.2$ flux ratios (see tab.~\ref{tab:merged_sensitivity_flux_upper}).
The results on the \Kold model can be compared with the previous ANTARES analysis with a shower and track sample extending from 2007 to 2015 \cite{albert:hal-01621706}. The current sensitivity, at $0.7$ flux ratio, is two times better than the previous one, at $1.4$ flux ratio. In the previous analysis, an upper limit of $1.1$ flux ratio was derived, to be compared with the current one of $1.4$ flux ratio. This is explained by an increase of the significance with the current dataset, going from $0.67 \sigma$ to the current significance of $1.28 \sigma$.

The different upper limits are also shown on fig.~\ref{fig:ANTARES_UL}. Each is rendered with a transparent-gradient procedure along energy axis, to show where the ANTARES detector could have seen the predicted signal. Most of the signal would be located between 100 GeV and a few TeV. One has to be careful in interpreting this plot: the signal/background weight of events increases with energy. Hence, the region above the TeV might impact the obtained constraints even if the statistics is lower for this regime.
Due to the low amount of signal events predicted (see tab.~\ref{tab:ns}), and fitted (see tab.~\ref{tab:merged_sensitivity_flux_upper}), the contribution of a few events might impact significantly the results, which would consequently change the upper limits.

\begin{figure}[ht]
    \centering
    \includegraphics[width=\linewidth]{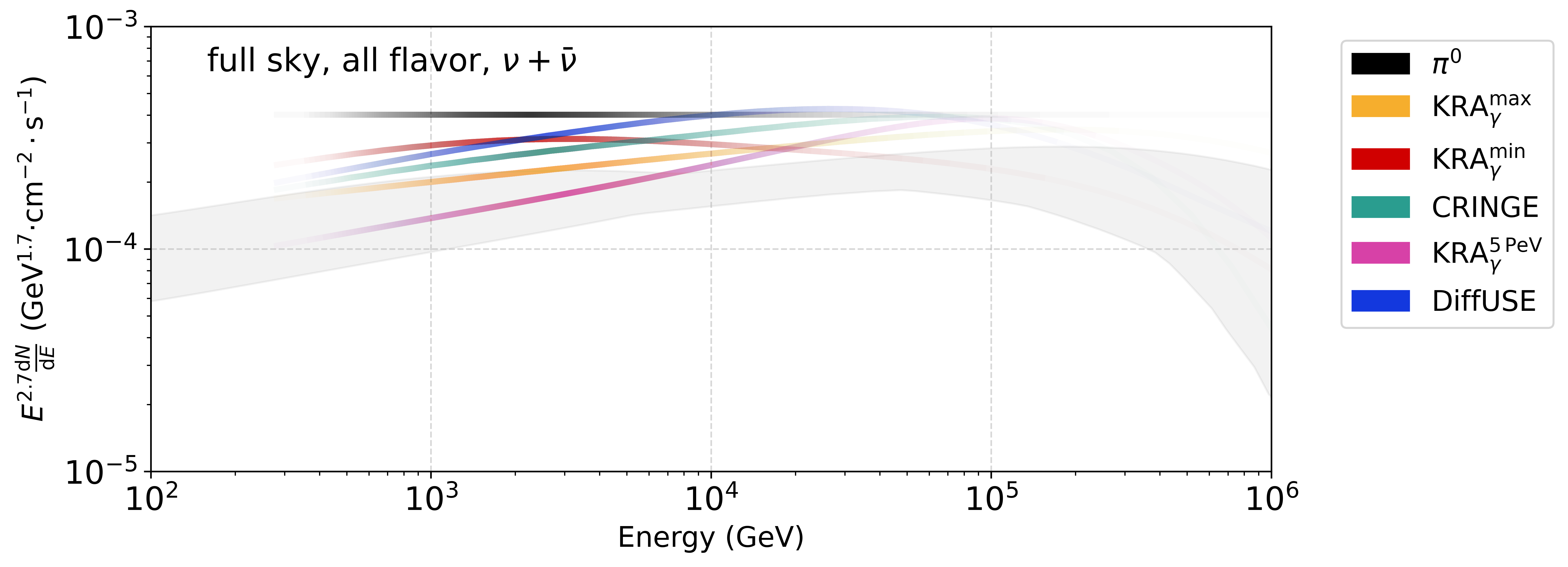}
    \caption{Comparison of the ANTARES upper limits (solid lines) at $90\%$ C.L.. The grey band represents the area covered by the flux predicted by the different models (except $\pi^0$). The upper limits are plotted according to a transparent gradient, to highlight the energy range of the ANTARES detector (hundreds of GeVs to a few TeVs).}
    
    \label{fig:ANTARES_UL}

    \includegraphics[width=\linewidth]{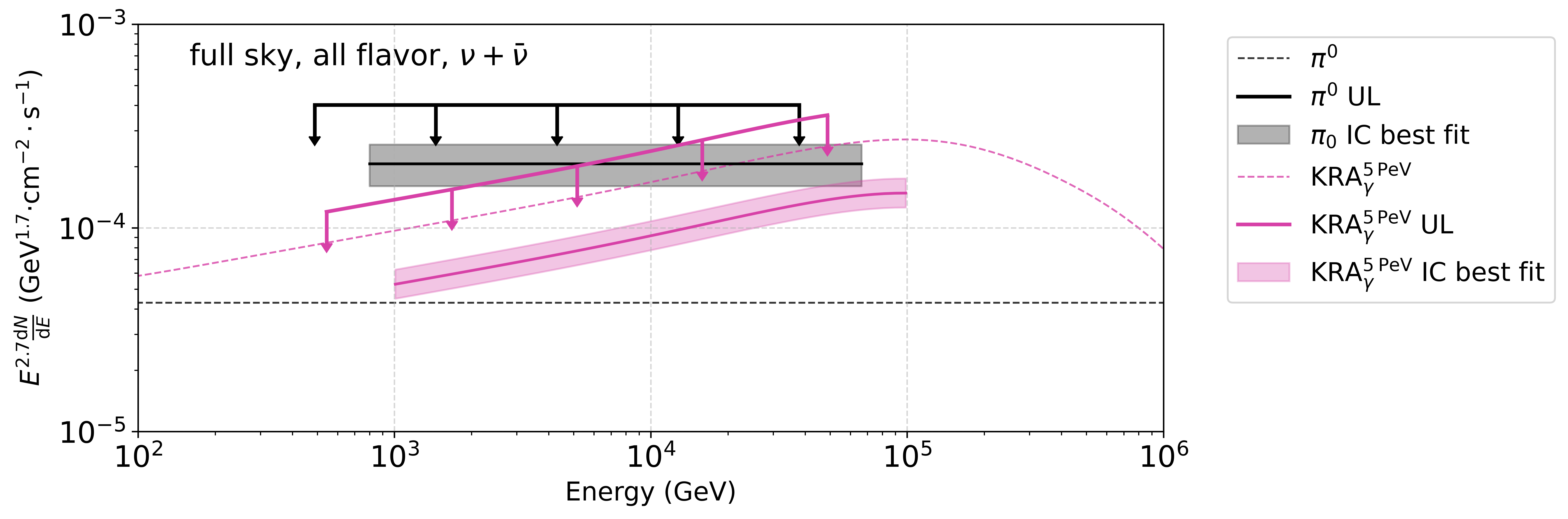}
    \caption{Comparison of the ANTARES upper limits at $90\%$ C.L. (see tab.~\ref{tab:merged_sensitivity_flux_upper}) with IceCube best fits for $\pi^0$ model (black) and \Kold  model (pink). The ANTARES constraints are compatible with IceCube results. The flux predicted by the models is represented with dashed lines.}
    \label{fig:IC-ANTARES}
\end{figure}

A comparison with IceCube results \cite{ICScience2023} shows no tension with respect to the \Kold  and $\pi^0$ best fits of IceCube, see fig.~\ref{fig:IC-ANTARES}. 
Instead of applying a color gradient as in fig.~\ref{fig:ANTARES_UL}, the solid line is extended to cover the central range in energy where $90\%$ of the signal is expected, as done in~\cite{ICScience2023}, and upper-limits are clearly above IceCube best fits.

The distribution of events in the region $|\ell|<30^\circ$ and $|b|<5^\circ$ is shown in fig.~\ref{fig:real-events} (upper plot) with the prediction from the signal PDF provided by the \Kmax model. It contains one track event which lies at the edge of the central peak (in black) of the \Kmax PDF. Even if the PDF is extended along the Galactic latitude $b$ for shower-like events, they all lie outside the central region. Convolving flatter templates with the ANTARES response yields an asymmetric profile in Galactic longitude, with an excess at negative longitudes. In the DiffUSE model, the predicted signal in the region $\ell \approx -40^\circ$ and $-10^\circ$ exceeds that expected near the Galactic Center (see fig.~\ref{fig:real-events}, bottom panel).

\begin{figure}[!ht]
    \centering
    \includegraphics[width=\linewidth]{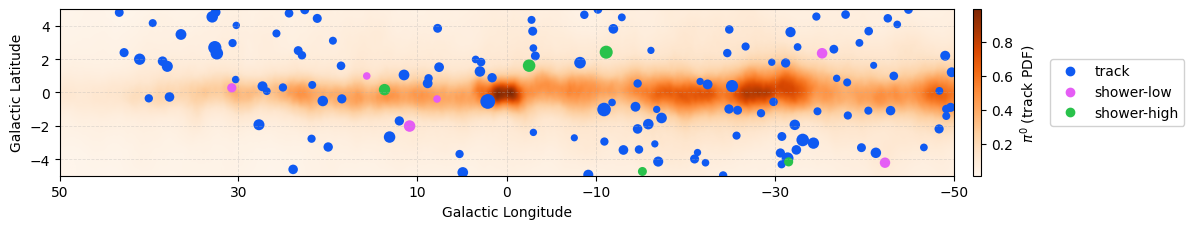}
    \includegraphics[width=\linewidth]{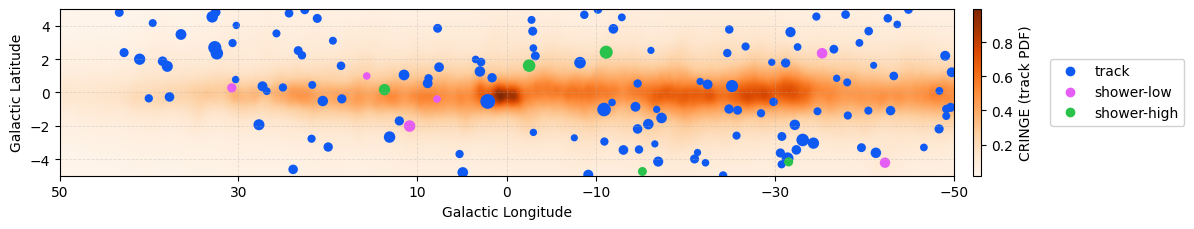}
    \includegraphics[width=\linewidth]{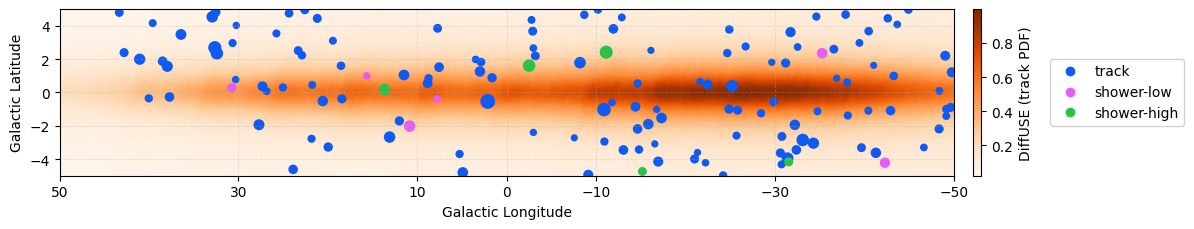}
    \includegraphics[width=\linewidth]{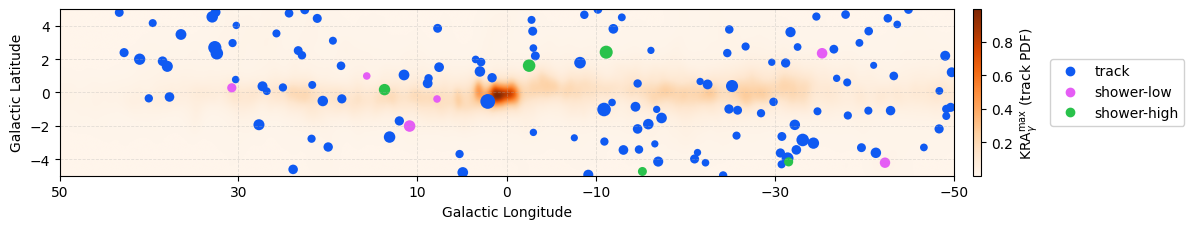}
    \includegraphics[width=\linewidth]{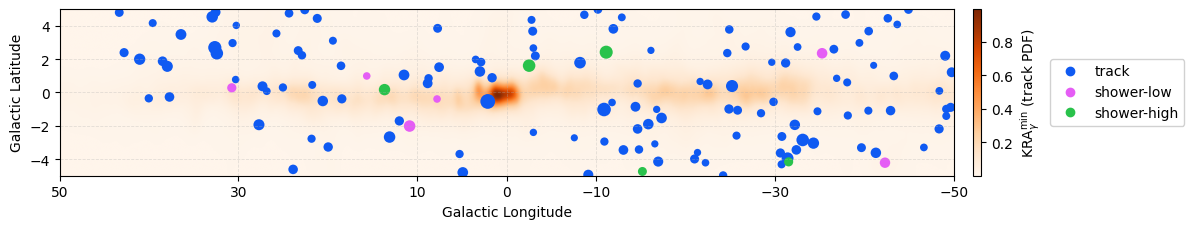}
    \includegraphics[width=\linewidth]{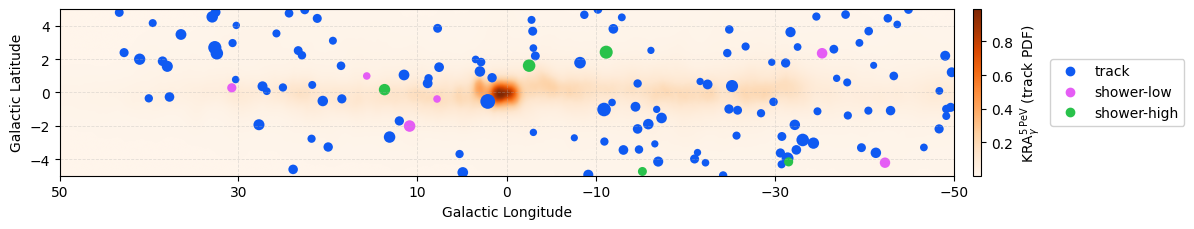}

    \caption{The distribution of real events (color dots) in the Galactic Center. The size of the dots is proportional to $(\log_{10}E)^{2.7}$. The distribution of real events is plotted on top of the signal PDF for the track channel integrated on energy, corresponding to the $\pi^0$, the CRINGE, the DiffUSE, the KRA$_\gamma^\mathrm{max}$, the KRA$_\gamma^\mathrm{min}$, and the KRA$_\gamma^\text{5  PeV}$ models (from top to bottom). In spite of the phenomenological templates being nearly symmetrical along the Galactic longitude, see fig.~\ref{fig:flux_l}, the signal ends up being higher at negative longitudes due to ANTARES response, which is more obvious for models whose distribution is flatter near the Galactic Center.}
    \label{fig:real-events}
\end{figure}

\subsection{The Galactic Ridge analysis}\label{GalRidge}

The Galactic Ridge is a region defined around the Galactic Center in Galactic coordinates by $|\ell| < 30^\circ$ and $|b| < 2^\circ$ where some excess in the neutrino flux is expected to be found.
A counting analysis has been performed to look for an excess on the Galactic Ridge, following the same on-off method as in the previous analysis \cite{GalRidge} that used data collected between May 2007 and February 2020 for track events, and until December 2020 for shower events. For the track-like channel, an excess of $1.9\sigma$ has been found in the Ridge. Due to the change in the samples, with two added years of data and new optimization cuts (see \ref{datasamples}), a similar on-and-off check gives consistent results (see tab.~\ref{tab:small_ridge}), with a p-value of $1.9\sigma$ for the track channel, which therefore confirms the presence of an excess in the Ridge with these new samples. These results on the Ridge provide a first model independent indication of Galactic neutrinos.

\begin{table}[h!]
\centering
\begin{tabular}{|l|c|c|c|c|}
\toprule
\hline
Category & $N_{\text{obs}}$ & $N_{\text{exp}}$ & Excess & p-value\\
\hline
Tracks & 35 & 26.1 & 8.9 & 5.5$\times10^{-2}$  ($1.9\sigma$)  \\
\hline
Showers (Low+High) & 9 & 8.1 & 0.9 & 0.42 ($0.80\sigma$)  \\
\hline
Total Combined & 44 & 34.2 & 9.8 & 6.1$\times10^{-2}$ ($1.9\sigma$)  \\
\hline
\end{tabular}
\caption{Number of observed and expected events in the Ridge. The estimated excess in the Ridge with a given two-tailed p-value for all the channels. A Poissonian counting method has been applied to obtain the results of this table, using the track, shower-low and shower-high samples described in this paper, and used for the template analysis.}
\label{tab:small_ridge}
\end{table}

\section{Conclusion}\label{Conclusion}
The full and final ANTARES upgoing neutrino sample with selective cuts allowing for a low atmospheric muon background have been tested for different phenomenological models of Galactic diffuse neutrino flux. While the model-independent Galactic Ridge analysis with the full ANTARES dataset indicates a hint of Galactic signal at $1.9\sigma$, as already seen by \cite{GalRidge}, tests of phenomenological models are the current best approaches to better understand the physics behind Galactic neutrinos.  The PDFs used in an unbinned maximum likelihood framework allow us to preserve the spatial-energy correlation of the templates and convolve them with the detector response aiming at a more precise modeling of the different signal responses. Even if the low statistics of ANTARES dataset does not allow to reach any significant discovery, a small hint, going from $1.0$ to $1.3\sigma$ is observed for the KRA$_\gamma^{5 \: \text{PeV}}$, KRA$_\gamma^\mathrm{min}$ and KRA$_\gamma^\mathrm{max}$ models. This analysis method is promising for testing Galactic diffuse emission models with larger datasets, particularly the ones from KM3NeT.

\section{Acknowledgments}

The authors acknowledge the financial support of the funding agencies:
Centre National de la Recherche Scientifique (CNRS), Commissariat \`a
l'\'ener\-gie atomique et aux \'energies alternatives (CEA),
Commission Europ\'eenne (FEDER fund and Marie Curie Program),
LabEx UnivEarthS (ANR-10-LABX-0023 and ANR-18-IDEX-0001),
R\'egion Alsace (contrat CPER), R\'egion Provence-Alpes-C\^ote d'Azur,
D\'e\-par\-tement du Var and Ville de La
Seyne-sur-Mer, France;
Bundesministerium f\"ur Bildung und Forschung
(BMBF), Germany; 
Istituto Nazionale di Fisica Nucleare (INFN), Italy;
Nederlandse organisatie voor Wetenschappelijk Onderzoek (NWO), the Netherlands;
Ministry of Education and Scientific Research, Romania;
MCIN for PID2021-124591NB-C41, -C42, -C43 and PDC2023-145913-I00 funded by MCIN/AEI/10.13039/501100011033 and by “ERDF A way of making Europe”, for ASFAE/2022/014 and ASFAE/2022 /023 with funding from the EU NextGenerationEU (PRTR-C17.I01) and Generalitat Valenciana, for Grant AST22\_6.2 with funding from Consejer\'{\i}a de Universidad, Investigaci\'on e Innovaci\'on and Gobierno de Espa\~na and European Union - NextGenerationEU, for CSIC-INFRA23013 and for CNS2023-144099, Generalitat Valenciana for CIDEGENT/2020/049, CIDEGENT/2021/23, CIDEIG/2023/20, ESGENT2024/24, CIPROM/2023/51, GRISOLIAP/2021/192 and INNVA1/2024/110 (IVACE+i), Spain;
Ministry of Higher Education, Scientific Research and Innovation, Morocco, and the Arab Fund for Economic and Social Development, Kuwait;

We also acknowledge the technical support of Ifremer, AIM and Foselev Marine
for the sea operation and the CC-IN2P3 for the computing facilities.
Vittoria Vecchiotti acknowledges support from European Research Council (ERC) under the ERC-2020-COG ERC Consolidator Grant (Grant agreement No.101002352) and the European Union’s Horizon Europe research and innovation programme under the Marie Skłodowska-Curie grant agreement No. 101208655 (CORNO GRANDE – COnstRaiNing the Origin of Galactic cosmic RAys using gamma-ray and Neutrino Diffuse Emissions).
V. Vecchiotti, G. Pagliaroli and F.L. Villante aknowledge support from Research grant number 2022E2J4RK ``PANTHEON: Perspectives in Astroparticle and Neutrino THEory with Old and New messengers" , PRIN 2022,  funded by the Italian Ministero dell’Universit\`a e della Ricerca (MUR) and by the European Union – Next Generation EU.

\bibliography{template}

\end{document}